\documentclass[aps,reprint,twocolumn,a4paper,superscriptaddress]{revtex4-2}

\usepackage{graphicx}
\usepackage{dcolumn}
\usepackage{bm}
\usepackage{setspace}
\usepackage{amsmath}

\begin{document}

\singlespacing

\title{Terahertz electro-optic effect in Bi$_2$Se$_3$ crystals}

\author{A. A. Melnikov}
\email{melnikov@isan.troitsk.ru}
\affiliation {Institute for Spectroscopy RAS, Troitsk, Moscow, 108840 Russia}
\affiliation {HSE University, Moscow, 109028 Russia}
\author{A. A. Sokolik}
\affiliation {Institute for Spectroscopy RAS, Troitsk, Moscow, 108840 Russia}
\affiliation {HSE University, Moscow, 109028 Russia}
\author{Yu. G. Selivanov}
\affiliation {P. N. Lebedev Physical Institute RAS, Moscow, 119991 Russia}
\author{S. V. Chekalin}
\affiliation {Institute for Spectroscopy RAS, Troitsk, Moscow, 108840 Russia}

\begin{abstract}

We report the observation of the electro-optic effect in Bi$_2$Se$_3$ crystals induced by an intense single-cycle terahertz pulse. The effect reveals itself as a transient change of the polarization state of a femtosecond laser pulse reflected from the crystal that is exposed to the terahertz electric field. The corresponding experimental signal follows the field with a sub-100-fs delay and can be represented as a linear combination of the terahertz electric field and its square. The linear and quadratic components are of comparable magnitude. The latter is almost independent of crystal orientation, while the former demonstrates three-fold rotational symmetry in agreement with the trigonal symmetry of the crystal surface. The electro-optic effect vanishes upon phase transition to the non-topological metal state induced by indium doping and also can be quenched by a femtosecond pre-pulse. We associate this effect with surface Dirac electronic states of Bi$_2$Se$_3$ and discuss its possible mechanisms.

\end{abstract}

\maketitle

\section{Introduction}

Bi$_2$Se$_3$ is often considered a model 3D topological insulator (TI) \cite{Hasan, Qi, Heremans, Yonezawa}. As-grown crystals have relatively high carrier concentration of up to $\sim$ 10$^{19}$ cm$^{-3}$. However, with the help of doping and electrostatic gating \cite{Kim, Checkelsky, Hong} or composition tuning \cite{Kushwaha} it is possible to reduce bulk conductivity and place the Fermi level in the vicinity of the Dirac point, bringing the crystal close to the ideal 3D TI. Surface electronic states of Bi$_2$Se$_3$ are characterized by the linear Dirac-like dispersion with spin helicity. The spin-momentum locking protects surface charge carriers from backscattering on nonmagnetic impurities and enables spin-polarized surface currents \cite{Ando}.

A broad research topic is the coupling of electromagnetic waves to surface states in TIs. Light can be used to probe fundamental microscopic processes associated with Dirac electrons and to control their motion. Two important counterparts in this context are the photogalvanic effect and the electro-optic effect. The former, namely, the generation of electron currents under the action of light, has been actively studied in Bi$_2$Se$_3$ crystals for more than a decade in view of potential applications in spintronics and optoelectronics \cite{Hosur, McIver, Kastl, Braun, Besbas, Seifert, Plank, Soifer, Connelly}. The most prominent and specific phenomenon here is the circular photogalvanic effect, when surface spin-polarized currents are induced via oblique illumination by circularly polarized radiation due to non-symmetric optical transitions at two sides of the Dirac cone \cite{Hosur, McIver, Kastl, Besbas, Seifert, Soifer}.

In turn, the electro-optic effect as a tool for studies of electronic states with nontrivial topology only recently gained considerable attention \cite{Cheng, Li, Ovalle, Morgado, deSousa, Ma}. In contrast to the standard configuration, in which a static electric field modifies the refractive index of wide-bandgap semiconductors or insulators, in this case an “intraband” version of the effect is primarily considered (cf. the nonlinear Hall effect \cite{Sodemann, Du, He}). Field-induced anisotropic changes of conductivity contain information on topological properties, such as Berry curvature, and on specific scattering processes.

Since application of high static or low-frequency electric fields to conductive samples can be problematic, it could be advantageous to study electro-optic effects at higher frequencies and for shorter durations of the waveforms, so that the effects of Joule heating are not destructive, while the relevant electronic processes have sufficient time to occur.

In the present work, we report the observation of the electro-optic effect in bulk crystals of Bi$_2$Se$_3$ using picosecond single-cycle terahertz pulses. The effect reveals itself as a transient rotation of polarization of the femtosecond probe pulse. The signal has two components of comparable magnitude, one of which is linear and the other is quadratic in the applied electric field. The amplitude of the linear component demonstrates the three-fold symmetry upon crystal rotation, in line with the $C_{3v}$ symmetry of the Bi$_2$Se$_3$ surface. At the same time, the effect vanishes upon transition to the topologically trivial phase induced by indium doping. These properties allow us to associate the observed electro-optic effect exclusively with the surface Dirac electrons of Bi$_2$Se$_3$. Though the effect is detected by probing interband optical transitions, its likely intraband origin is indicated by the fact that it can be quenched by a preceding femtosecond pulse that heats the electronic ensemble smearing the Fermi distribution.

\section{Experimental details}

The crystals of (Bi$_{1-x}$In$_x$)$_2$Se$_3$ with $x$ = 0, 0.025, 0.05, and 0.1 were grown by the modified Bridgman method \cite{Zhukova} and cleaved along the (0001) basal plane. Experiments were performed using the standard pump-probe layout. Pump terahertz pulses were generated via optical rectification of femtosecond laser pulses with tilted fronts in a crystal of lithium niobate \cite{Stepanov}. For that purpose, a fraction of the output beam from a regenerative Ti:sapphire amplifier (Spitfire Pro XP, Spectra Physics) was used (1.8 mJ per pulse at 800 nm, duration of 80 fs, 1 kHz repetition rate). With the help of parabolic mirrors, the terahertz beam was collimated and focused onto the sample so that the peak electric field of the resultant $\sim$ 1 ps terahertz pulses was up to $\sim$ 500 kV/cm, while their peak frequency was near 1 THz. Pump femtosecond pulses at wavelengths of 1300 and 650 nm were the idler of an optical parametric amplifier (TOPAS, Light Conversion) and its second harmonic, respectively, with a duration of $\sim$ 80 fs.

\begin{figure}
\begin{center}
\includegraphics[width=0.7\columnwidth]{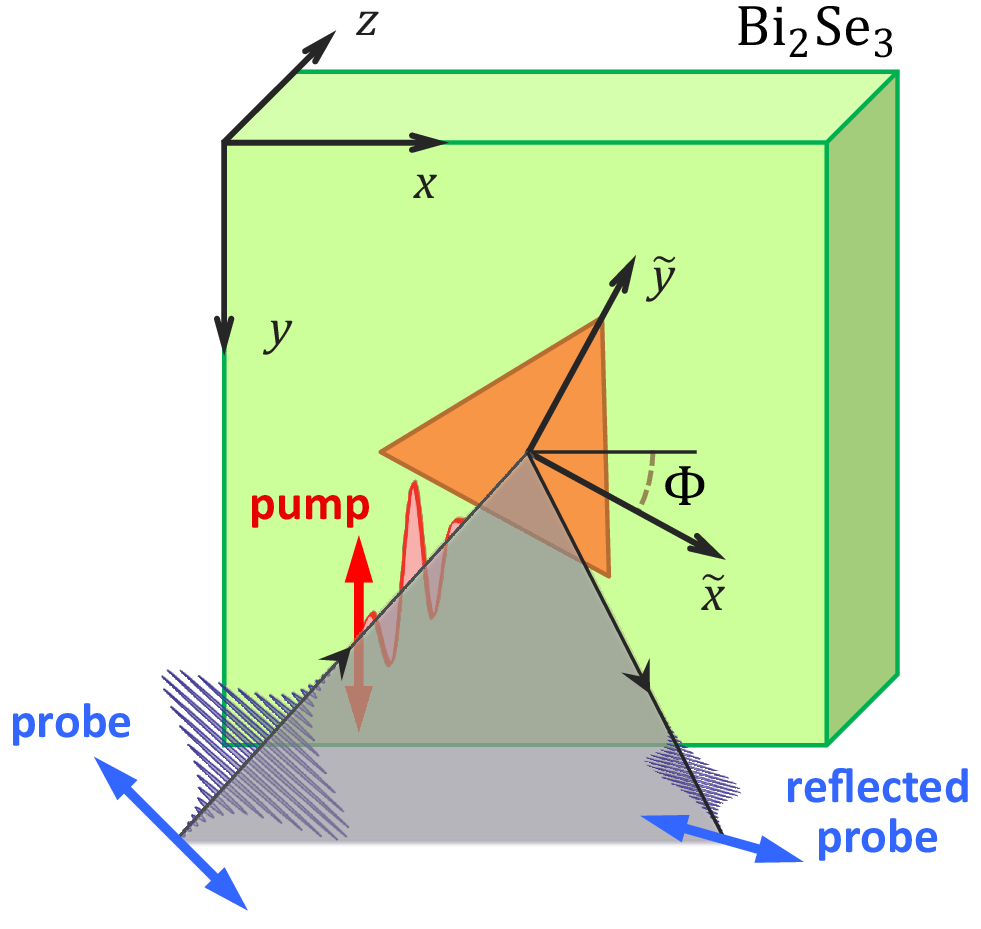}
\end{center}
\caption{Reflection of the probe pulse (blue), which is polarized at $45^\circ$ with respect to the plane of incidence (gray). The terahertz pump pulse (red) is $s$-polarized. $\Phi$ is the crystal rotation angle (the crystal is rotated in the surface plane).}
\end{figure}

In order to detect the electro-optic effect induced by the terahertz field, we used weak femtosecond probe pulses at 800 nm. The pump and probe beams were incident onto the sample at an angle $\theta \sim 8^\circ$, while the initial polarization of the probe pulse was set to 45$^\circ$ relative to the vertical polarization of the pump terahertz pulses (see Fig. 1). Intensities of the vertical and horizontal polarization components of the reflected probe pulses $I_y$ and $I_x$ were detected using a Wollaston prism and a pair of amplified photodiodes. The measurement was repeated multiple times for open and closed pump beams (modulated by an optical chopper) and the normalized difference signal $S=1-(I^*_y/I^*_x)/(I_y/I_x)$  was calculated and then averaged (here the asterisk indicates the intensities measured with the opened pump beam). For small angles of incidence $S \approx - 4\varphi$, where $\varphi$ is the rotation of polarization of the reflected probe pulse. In order to measure pump-induced changes of reflectivity, a similar procedure was performed, in which intensities of the reflected probe beam and of the reference beam were measured instead of intensities of the polarization components. 

\section{Results}

Figure 2 illustrates transient changes of reflectivity of the (Bi$_{1-x}$In$_x$)$_2$Se$_3$ crystal samples detected at 800 nm and induced by the pump terahertz pulse or by the femtosecond pulse with the central wavelength of 1300 nm. We consider terahertz pulses to be too weak to cause efficient multiphoton or tunneling transitions between different electronic bands of Bi$_2$Se$_3$. Therefore, their main initial effect is Joule heating of bulk electrons near the Fermi level. This is evidenced in part by the similarity of the $\sim$ 500 fs rise time of the signal, and the rise time of the integrated squared terahertz waveform (Fig. 2(b), note the much faster rise of the transient induced by the 1300 nm 80 fs pump pulse). 

\begin{figure}
\begin{center}
\includegraphics{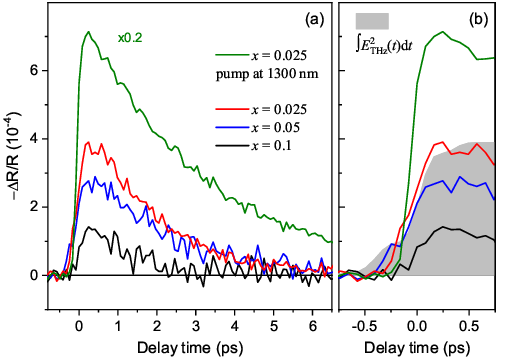}
\end{center}
\caption{(a) Transient changes of reflectivity of (Bi$_{1-x}$In$_x$)$_2$Se$_3$ crystals with various indium content $x$ induced by terahertz pulses or by femtosecond laser pulses with a central wavelength of 1300 nm. (b) Magnified view of the initial rise of the traces shown in panel (a). The shaded area corresponds to the integrated squared electric field of the terahertz pulse.}
\end{figure}

Relaxation of the terahertz pump-induced changes of reflectivity has a characteristic time of $\tau$ = 1.8 ps, which is comparable to the $\tau$ = 3.2 ps observed in the case of excitation by femtosecond pulses at 1300 nm. These values are in general agreement with the characteristic cooling times of bulk electrons in Bi$_2$Se$_3$ crystals obtained in time-resolved electron photoemission studies (see e. g. \cite{Sobota, Crepaldi, Ponzoni}). It should be noted that the increase of In content leads to the decrease of the signal induced by the terahertz pulse by a factor of $\sim$ 2.7 for $x$ = 0.1 (see Fig. 2(a)). This can be explained by a certain decrease of carrier concentration that occurs as a result of doping by indium atoms.

\begin{figure}
\begin{center}
\includegraphics{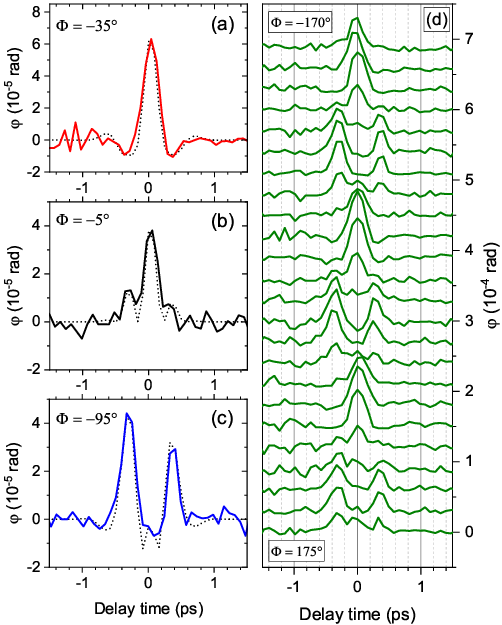}
\end{center}
\caption{ (a)--(c) Polarization rotation $\varphi(t)$ of the probe pulse induced by the terahertz pulse for the (Bi$_{1-x}$In$_x$)$_2$Se$_3$ crystal with $x$ = 0.025 for three specific sample orientations (see text). Dotted lines indicate fitting curves. (d) Evolution of the $\varphi(t)$ waveform upon crystal rotation. The curves are shifted vertically for clarity.}
\end{figure}

Next, we proceed to the description of the observed electro-optic effect. Transient polarization rotation of the probe pulse $\varphi(t)$ induced by the terahertz electric field in the crystal sample with $x$ = 0.025 is plotted in Fig. 3 for a number of crystal orientations. The corresponding signals for the Bi$_2$Se$_3$ crystal ($x$ = 0) are similar and will not be shown here. We note that the electro-optic effect was observed for freshly cleaved crystals (after $\sim$ 10 min from cleavage), as well as for crystals that were exposed to ambient air for at least several days. In Fig. 3(d) one can see periodic alternation of several characteristic waveforms upon crystal rotation. We have found that this signal can be relatively well fitted by a linear combination $\varphi(t) = A_1E(t) + A_2E^2(t)$, where $E(t)$ is the terahertz electric field, and $A_2 > 0$. Three typical shapes of $\varphi(t)$ detected for those rotation angles, for which $A_1\sim A_2$, $A_1 = 0$, and $A_1\sim - A_2$, are shown in Fig. 3 (a), (b), and (c), respectively.

\begin{figure}
\begin{center}
\includegraphics{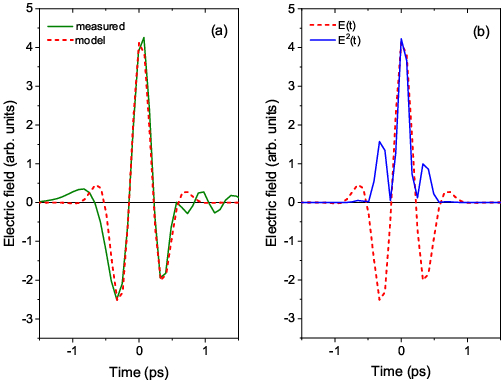}
\end{center}
\caption{(a) Measured time dependence of the electric field of the terahertz pulse (green solid line) and the model electric field (red dashed line). (b) Model electric field (red dashed line) and its normalized square (blue solid line).}
\end{figure}

It is necessary to note that in order to achieve a better quality of the fit we have used a specific model waveform $E(t)$ instead of the measured terahertz field. It is determined by the formula $E(t) \propto e^{-at^2}\cos(bt+\phi)$ and is compared with the terahertz waveform in Fig. 4(a). It can be seen that the waveforms are rather similar except for the region of the first major negative peak near $-0.5$ ps. A possible reason for this discrepancy can be a difference between the measured terahertz field and the real electric field acting on the crystal. In order to attenuate terahertz pulses incident onto the ZnTe crystal for the electro-optic detection, we used a 1 mm thick fused silica plate covered with a thin metal layer. High frequency absorption in this plate could cause distortion of the terahertz pulse shape.

The values of $A_1$ and $A_2$ that were obtained as a result of the least squares fitting of the waveforms shown in Fig. 3 are plotted in Fig. 5. It can be seen that the linear component is a combination of $\sin (3\Phi + \phi)$ and a constant offset, and thus demonstrates the three-fold rotation symmetry. The second component of the signal that is quadratic in the electric field is almost independent of crystal orientation with a small admixture of the $3\Phi$ harmonic.

\begin{figure}
\begin{center}
\includegraphics{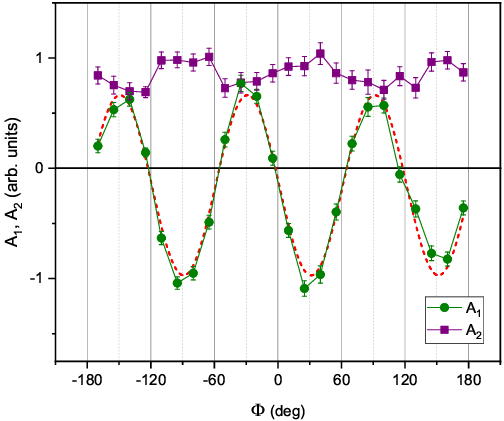}
\end{center}
\caption{Amplitudes of the linear ($A_1$) and quadratic ($A_2$) components obtained by least squares fitting of measured $\varphi(t)$ waveforms for various crystal orientations. The dashed line indicates a $\sin (3\Phi + \phi)$ fit.}
\end{figure}

For peak electric fields, the strengths of which are related as 1:2, the amplitudes $A_1$ and $A_2$ in the corresponding $\varphi(t)$ waveforms should be related as 1:2 and 1:4, respectively. In order to check this property, we have measured the signal $\varphi(t)$ for peak electric fields of the pump terahertz pulse as large as 250 kV/cm and 500 kV/cm. The results are shown in Fig. 6. The fitting procedure provided the relations 1:2.2$\pm$0.4 and 1:3.2$\pm$0.5 for the amplitudes $A_1$ and $A_2$ at lower and higher peak electric fields. The agreement with the expected value is good for the linear component and somewhat worse for the quadratic component. 

\begin{figure}
\begin{center}
\includegraphics{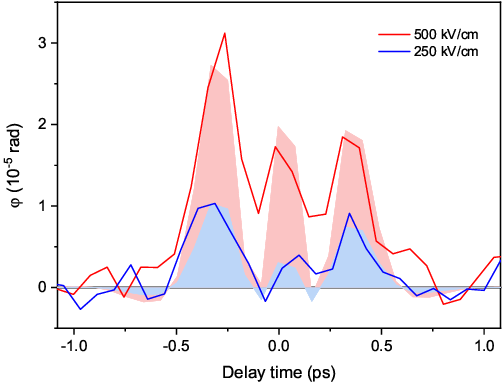}
\end{center}
\caption{Waveforms $\varphi(t)$ measured using peak electric fields of 500 kV/cm (red line) and 250 kV/cm (blue line) and the fitting curves (indicated by shaded areas of corresponding colors).}
\end{figure}

Next, we compare the electro-optic effect for (Bi$_{1-x}$In$_x$)$_2$Se$_3$ crystals with various indium content $x$. It is known that for the (Bi$_{1-x}$In$_x$)$_2$Se$_3$ system the increase of indium content causes a transition to the topologically trivial phase at $x\sim$ 6--7 $\%$ \cite{Brahlek, Wu, Lou, Sánchez-Barriga}. The substitution of Bi atoms by lighter In atoms effectively decreases spin-orbit coupling and thereby relaxes band inversion, which is required for the formation of the topological surface states. With increasing indium content, the bulk band gap closes and opens again but without band inversion. The resulting crystal is a trivial metal without topological surface states. Figure 7 contains three sets of $\varphi(t)$ transients measured for crystals with $x$ = 0.025, 0.05, and 0.1 for a number of crystal rotation angles. It can be seen that the electro-optic effect vanishes for $x$ = 0.1, at which the crystal is a non-topological metal (or at least its magnitude drops below the noise level).

The residual oscillations that are visible after zero delay time in Fig. 7(c) are due to coherent $E_u^1$ and $E_g^2$ phonons generated by the terahertz pulse (see the inset for their averaged spectrum). We note that the amplitude of coherent phonons that can be excited in epitaxial Bi$_2$Se$_3$ films by terahertz pulses with similar peak field strengths is almost two orders of magnitude higher \cite{Melnikov1, Melnikov2}. One possible explanation of this fact is that in the latter case the anisotropy of transmittance is detected, while modulation of reflectivity by coherent phonons of the same amplitude is expected to be considerably weaker. Another factor that can decrease the efficiency of coherent phonon generation in Bi$_2$Se$_3$ (in particular, of the $E_g^2$ mode) is the screening of the resonant infrared active $E_u^1$ mode by free electrons. Indeed, as concentration of the electrons is reduced by indium doping in our case, the oscillations become more pronounced.

\begin{figure}
\begin{center}
\includegraphics{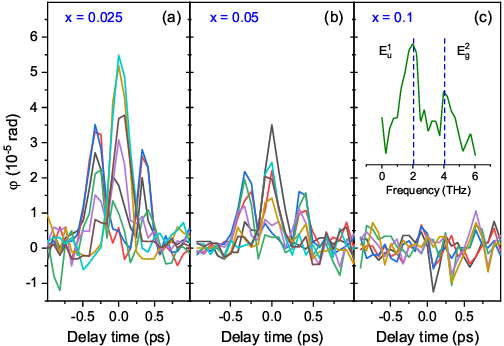}
\end{center}
\caption{Sets of $\varphi(t)$ waveforms measured for several selected $\Phi$ for (Bi$_{1-x}$In$_x$)$_2$Se$_3$ crystals with $x$ = 0.025 (a), $x$ = 0.05 (b), and $x$ = 0.1 (c). The inset to panel (c) shows averaged spectrum of the oscillations visible after zero delay time.}
\end{figure}

Another notable property of the observed electro-optic effect, that we have discovered, is its high sensitivity to the preliminary excitation by an additional femtosecond laser pulse at a central wavelength in the visible or near infrared range. We have found that it is possible to quench the transient response $\varphi(t)$ using such a pre-pulse, so that its magnitude decreases below the noise level. This effect is illustrated in Fig. 8 for $\lambda_\mathrm{prepulse} = 650$ nm. We have selected a certain crystal orientation, at which the linear and the quadratic components of the signal $\varphi(t)$ have positive and approximately equal amplitudes $A_1 \approx A_2$ (similar to the waveform shown in Fig. 3(a)). It can be seen that the peak amplitude of the $\varphi(t)$ waveform is close to zero if the femtosecond pre-pulse acts immediately before the terahertz pump pulse. As the delay time between these two pulses is increased, the signal is restored to its original form detected without preliminary excitation. The dependence of the peak amplitude on the delay time in Fig. 8(b) is not exponential. However, it is possible to assign to it a “characteristic time” of roughly 3 ps, which is similar to the timescale of electronic cooling, as discussed above.

\begin{figure}
\begin{center}
\includegraphics{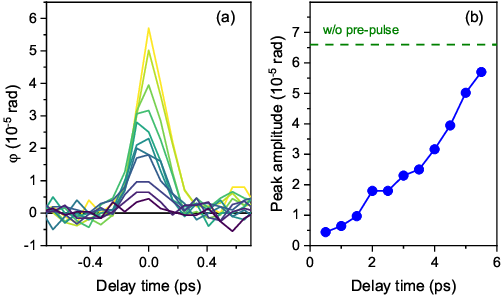}
\end{center}
\caption{(a) $\varphi(t)$ signals measured for various temporal delays of the femtosecond laser pre-pulse. Measurements were performed for the crystal with $x$ = 0.025 and at a certain $\Phi$, for which $A_1 \approx A_2$. Minimum and maximum delays correspond to violet and yellow curves, respectively. (b) The dependence of the peak amplitude of $\varphi(t)$ on the delay time between the pre-pulse and the terahertz pulse. The minimum delay corresponds to the arrival of the pre-pulse at $t \sim -1$ ps in Fig. 4(a). The dashed horizontal line indicates the peak amplitude obtained without the pre-pulse.}
\end{figure}

It is also interesting to follow the dependence of $\varphi(t)$ on the intensity of the femtosecond pre-pulse. As can be seen in Fig. 9, the peak amplitude of $\varphi(t)$ is roughly inversely proportional to the fluence of the pre-pulse, and the electro-optic effect can be effectively quenched already by pulses with moderate fluences of $\sim$ 0.1 mJ/cm$^2$.

\begin{figure}
\begin{center}
\includegraphics{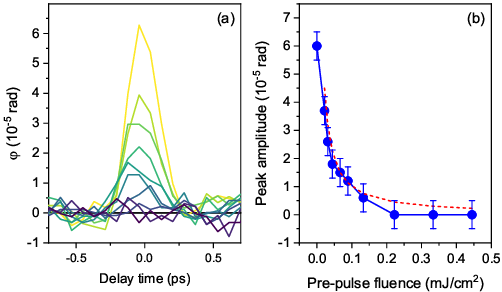}
\end{center}
\caption{(a) $\varphi(t)$ signals measured for various fluences of the femtosecond laser pre-pulse. Measurements were performed for the crystal with $x$ = 0.025 and at a certain $\Phi$, for which $A_1 \approx A_2$. Minimum and maximum fluences correspond to yellow and violet curves, respectively. (b) The dependence of the peak amplitude of $\varphi(t)$ on the pre-pulse fluence $F$. The dashed line indicates the $\propto 1/F$ fit.}
\end{figure}

\section{Symmetry analysis}

In order to complement the obtained experimental results by symmetry analysis, we consider reflection of the probe beam incident on the crystal with dielectric permittivity $\varepsilon$ at an angle $\theta$ (Fig. 1). The incident probe pulse is polarized at $45^\circ$ with respect to the plane of incidence, so $p$- and $s$-polarized components of its electric field are equal: $E_\mathrm{ip}=E_\mathrm{is}=E_\mathrm{i}$. The corresponding components of the electric field of the reflected wave are $E_\mathrm{rp}=E_\mathrm{i}r_\mathrm{p}$ and $E_\mathrm{rs}=E_\mathrm{i}r_\mathrm{s}$, with the amplitude reflection coefficients $r_\mathrm{p,s}$ given by Fresnel formulas.
The electric field of the pump terahertz pulse acting on the crystal sample causes small increments $\delta E_\mathrm{rp}$ and $\delta E_\mathrm{rs}$ in the polarization components of the reflected probe pulse. This change of the polarization state causes a non-zero signal
\begin{equation}
S=1-\frac{|E_\mathrm{rs}+\delta E_\mathrm{rs}|^2/|E_\mathrm{rs}|^2}{|E_\mathrm{rp}+\delta E_\mathrm{rp}|^2/|E_\mathrm{rp}|^2}.
\end{equation}
As we have mentioned above, it can be shown that for small angles of incidence ($r_\mathrm{s} \approx - r_\mathrm{p}$) we actually detect polarization rotation of the probe pulse by the angle $\varphi \approx - S/4$. 

 The probe pulse induces electric fields inside the sample near its surface with the amplitudes $E_x=E_\mathrm{i}(1-r_\mathrm{p})\cos\theta$, $E_y=E_\mathrm{i}(1+r_\mathrm{s})$, $E_z=E_\mathrm{i}(1-r_\mathrm{p})\sin\theta/\varepsilon$. On the other hand, the $s$-polarized pump pulse induces slowly varying terahertz field $E_y^\mathrm{THz}$ at the sample surface. The combined action of the pump and probe pulses gives rise to the nonlinear response of surface currents
  
 \begin{align}
\delta j_\alpha & = \sum_{\beta=x,y,z}\delta\sigma_{\alpha\beta}E_\beta \nonumber\\
&= \sum_{\beta=x,y,z}[\chi^\mathrm{(1)}_{\alpha\beta y}E_y^\mathrm{THz}+\chi^\mathrm{(2)}_{\alpha\beta yy}(E_y^\mathrm{THz})^2]E_\beta,\label{delta_j}
\end{align}
where $\alpha=x,y$, $\sigma_{\alpha\beta}$ is the optical conductivity tensor, $\chi^\mathrm{(1)}_{\alpha\beta\gamma}=\chi^\mathrm{(1)}_{\alpha\beta\gamma}(\omega,\omega,0)$ is the third-rank tensor of the linear electro-optic effect, and $\chi^\mathrm{(2)}_{\alpha\beta\gamma\rho}=\chi^\mathrm{(2)}_{\alpha\beta\gamma\rho}(\omega,\omega,0,0)$ is the fourth-rank tensor of the quadratic electro-optic effect. Both characterize the combined response at an optical frequency $\omega$ to the optical and quasistatic terahertz fields. For bulk dielectric media, the linear- and quadratic electro-optic effects are usually referred to as Pockels and Kerr effects, respectively, and the changes of dielectric function $\varepsilon_{\alpha\beta}$ due to static electric field is characterized by the third-rank Pockels and fourth-rank Kerr effect tensors. Since the dielectric function is not suitable to quantify the response of quasi-two-dimensional surface states, we describe the nonlinear effects in Eq.~(\ref{delta_j}) in terms of transient changes $\delta\sigma_{\alpha\beta}$ of the optical conductivity tensor due to the terahertz field. Although $\sigma_{\alpha\beta}$ is $\pi/2$ phase-shifted with respect to equivalent $\varepsilon_{\alpha\beta}$, their symmetry properties should be the same, so $\chi^\mathrm{(1)}_{\alpha\beta\gamma}$ and $\chi^\mathrm{(2)}_{\alpha\beta\gamma\rho}$ have the same symmetry properties as, respectively, the Pockels and Kerr effect tensors.

We also emphasize that the instantaneous relation between $\delta j_\alpha$ and $E^\mathrm{THz}_y$ is assumed in Eq. (2) for simplicity of the analysis, although a certain retardation should be present here, as well as a possible time dependence of the $\chi^{(1)}_{\alpha\beta\gamma}$ and $\chi^{(2)}_{\alpha\beta\gamma\rho}$ tensors due to transient heating of the sample. However, the symmetry properties are universal with respect to these factors and would not have changed if we had added the retardation explicitly. In addition, the delay between the signal and the field is rather small (see the discussion below), and the transient increase of the electron temperature induced by the terahertz pulse is moderate (not larger than $\sim$ 200-300 K, according to our estimates).
 
Nonlinearly induced oscillating currents $\delta j_{x,y}$ generate additional contributions to the reflected wave, the amplitudes of which can be calculated using the Maxwell equations: $\delta E_\mathrm{rp}=(2\pi/c)(1-r_\mathrm{p})\delta j_x$, $\delta E_\mathrm{rs}=-(2\pi/c\cos\theta)(1+r_\mathrm{s})\delta j_y$. To the first order, it gives rise to the signal
\begin{equation}
S\approx\frac{4\pi}c\,\mathrm{Re}\left(\frac{1-r_\mathrm{p}}{r_\mathrm{p}}\frac{\delta j_x}{E_i}+\frac{1+r_\mathrm{s}}{r_\mathrm{s}\cos\theta}\frac{\delta j_y}{E_i}\right).
\end{equation}

In order to obtain the nonlinear current response (\ref{delta_j}), we need to transform the tensor $\chi_{\alpha\beta\gamma}$ from crystallographic coordinate system $(\tilde{x},\tilde{y},\tilde{z})$ to coordinate system $(x,y,z)$ locked to the plane of incidence (see Fig. 1). Given the $C_{3v}$ (or $3m$) symmetry of the Bi$_2$Se$_3$ surface, we obtain the following general form of the Pockels effect tensor
\begin{equation}
\chi^\mathrm{P}_{\tilde\alpha\tilde\beta\tilde\gamma}=\bordermatrix{     
     & \tilde{x} & \tilde{y} & \tilde{z} \cr
\tilde{x}\tilde{x} &  0 & -b & c  \cr
\tilde{y}\tilde{y} &  0 &  b & c  \cr
\tilde{z}\tilde{z} &  0 &  0 & d  \cr
\tilde{y}\tilde{z} &  0 &  a & 0  \cr
\tilde{z}\tilde{x} &  a &  0 & 0  \cr
\tilde{x}\tilde{y} & -b &  0 & 0
},
\end{equation}
where rows and columns correspond to $\tilde\alpha\tilde\beta$ and $\tilde\gamma$ indices, respectively \cite{Agullo-Lopez, Newnham}. The Kerr effect tensor $\chi^\mathrm{K}_{\tilde\alpha\tilde\beta\tilde\gamma\tilde\rho}$ for the same symmetry group is 
\begin{equation}
\bordermatrix{     
     & \tilde{x}\tilde{x} & \tilde{y}\tilde{y} & \tilde{z}\tilde{z} & \tilde{y}\tilde{z} & \tilde{z}\tilde{x} & \tilde{x}\tilde{y} \cr
\tilde{x}\tilde{x} & a' &  b' & c' &  f' & 0  & 0  \cr
\tilde{y}\tilde{y} & b' &  a' & c' & -f' & 0  & 0  \cr
\tilde{z}\tilde{z} & d' &  d' & e' &  0  & 0  & 0  \cr
\tilde{y}\tilde{z} & g' & -g' & 0  &  h' & 0  & 0  \cr
\tilde{z}\tilde{x} & 0  &  0  & 0  &  0  & h' & g' \cr
\tilde{x}\tilde{y} & 0  &  0  & 0  &  0  & f' & \frac{a'-b'}2
},
\end{equation}
where rows and columns correspond to $\tilde\alpha\tilde\beta$ and $\tilde\gamma\tilde\rho$ indices, respectively \cite{Agullo-Lopez, Newnham}. After rotation by the angle $\Phi$ around the $z=\tilde{z}$ axis (see Fig. 1), we obtain the tensors $\chi^\mathrm{P}_{\alpha\beta\gamma}$ and $\chi^\mathrm{K}_{\alpha\beta\gamma\rho}$ entering Eq.~(\ref{delta_j}). After straightforward calculations, we obtain the following result for the signal $S$:
\begin{align}
S&=E_y^\mathrm{THz}(A+B\cos3\Phi+C\sin3\Phi)\nonumber\\
&+(E_y^\mathrm{THz})^2(A'+B'\cos3\Phi+C'\sin3\Phi),
\end{align}
where 
\begin{align}
A&=\frac{4\pi}c\,\mathrm{Re}\left[a\frac{(1-r_\mathrm{p})(1+r_\mathrm{s})\tan\theta}{\varepsilon r_\mathrm{s}}\right],\\
B&=\frac{4\pi}c\,\mathrm{Re}\left[-b\frac{(1-r_\mathrm{p})^2\cos\theta}{r_\mathrm{p}}+b\frac{(1+r_\mathrm{s})^2}{r_\mathrm{s}\cos\theta}\right],\\
C&=\frac{4\pi}c\,\mathrm{Re}\left[b(1-r_\mathrm{p})(1+r_\mathrm{s})\left(\frac1{r_\mathrm{p}}+\frac1{r_\mathrm{s}}\right)\right]
\end{align}
are responsible for the linear electro-optic effect, and
\begin{align}
A'&=\frac{4\pi}c\,\mathrm{Re}\left[b'\frac{(1-r_\mathrm{p})^2\cos\theta}{r_\mathrm{p}}+a'\frac{(1+r_\mathrm{s})^2}{r_\mathrm{s}\cos\theta}\right],\\
B'&=\frac{4\pi}c\,\mathrm{Re}\left[-g'\frac{(1-r_\mathrm{p})(1+r_\mathrm{s})\tan\theta}{\varepsilon r_\mathrm{s}}\right],\\
C'&=\frac{4\pi}c\,\mathrm{Re}\left[-g'\frac{(1-r_\mathrm{p})^2\sin\theta}{\varepsilon r_\mathrm{s}}\right].
\end{align}
are responsible for the quadratic electro-optic effect.

In the experiments, we have detected the $3\Phi$-harmonic component and the isotropic component in the linear response, which correspond to the coefficients $A$, $B$, and $C$. The quadratic response to the terahertz pump pulse was dominated by the isotropic component, which corresponds to $A'$. It also contained a small contribution of the $3\Phi$-harmonic associated with the coefficient $g'$ of the quadratic electro-optic effect, which mixes in-plane ($\tilde{x},\tilde{y}$) and out-of-plane ($\tilde{z}$) electron motions.

\section{Discussion}

In an attempt to assign the detected electro-optic effect to certain microscopic processes, we first consider possible electronic states that can be involved in this phenomenon. Taking into account the observed three-fold rotational symmetry of the linear component and the disappearance of the signal for the topologically trivial phase, we suppose that the electro-optic effect is associated with surface Dirac electronic states. It should be noted that there are two other specific types of states that can in principle exist at the surface of Bi$_2$Se$_3$, namely, the trivial two-dimensional electron gas (2DEG) \cite{Bianchi, King} and Rashba-split 2DEG states \cite{King, Zhu}. As for the trivial 2DEG, we believe that its possible contribution to the electro-optic effect can be neglected, because both the TI and the topologically trivial phase of Bi$_2$Se$_3$ can in principle host 2DEG. The contribution of Rashba states cannot be excluded relying only on the dependence of $\varphi(t)$ on the indium content, since the introduction of lighter indium atoms is expected to reduce the strength of spin-orbit coupling. However, as can be inferred from the literature, the emergence of Rashba spin-split states on the surface of Bi$_2$Se$_3$ is not a universal phenomenon and was observed only in certain Bi$_2$Se$_3$ crystals, typically with a rather strong $n$-type doping (see e. g. supplementary information in Ref. \cite{McIver}). Therefore, we believe that the observed electro-optic effect is not related to possible 2DEG and Rashba spin-split electronic states.

The observed quenching of the electro-optic effect by the femtosecond pre-pulse suggests that the optical response to the terahertz electric field is determined mainly by the intraband electronic processes in the vicinity of the Fermi level. In this case, the detected anisotropy of reflectivity is the manifestation of the transient anisotropy of the electronic distribution. The latter is probed via optical transitions to higher lying or from lower lying electronic bands. Intuitively, the effect is stronger if the Fermi distribution is sharper.

The main effect of the additional pulse is heating the electron ensemble, which leads to smearing of the Fermi distribution of both the bulk and the surface electrons. If we follow the dependence of $\varphi(t)$ on the intensity of the femtosecond pre-pulse shown in Fig. 9, it can be seen that already the pulses with a moderate fluence of $\sim$ 0.1 mJ/cm$^2$ can effectively quench the electro-optic effect. Estimates of the maximum transient electronic temperature in Bi$_2$Se$_3$ that can be found in the literature for pump fluences of the same order are $\sim$ 1000 K \cite{Crepaldi, Wang, Ponzoni}. This value is much lower than the energy of the probe photons in our experiments (1.55 eV, $\sim$ 18000 K). Therefore, such a “local” change of the electronic distribution is unlikely to considerably affect the total probability of optical transitions between states separated by 1.55 eV, which would determine the interband electro-optic effect.

The amount of smearing of the Fermi distribution can be characterized by the maximum value of its derivative (at $E = \mu$, where $\mu$ is the chemical potential), which is inversely proportional to the temperature. Interestingly, we can also assume the inverse proportionality of $\varphi$ on the maximum temperature $T_\mathrm{el}$ of the electrons heated by the femtosecond pre-pulse. Indeed, as can be seen in Fig. 9(b), $\varphi \propto 1/F$, where $F$ is the pre-pulse fluence. At the same time, for metals $T_\mathrm{el} \propto F$. This surprising correspondence can be regarded as additional evidence for the intraband character of the detected electro-optic effect in Bi$_2$Se$_3$.

The most common intraband effect that occurs in a metal under the action of the electric field of a terahertz pulse is Joule heating of the electron ensemble. If the rate of the electron-electron scattering is sufficiently high, a thermalized electron distribution will evolve during the pulse, the temperature of which approximately follows the integral of the squared electric field (provided the characteristic relaxation time of the electronic temperature is rather long). In Bi$_2$Se$_3$ this process reveals itself as the transient changes of reflectivity shown in Fig. 2 and mainly associated with the heating of bulk electrons. These changes, however, are cancelled out almost completely in the process of detection of the $\varphi(t)$ signal. The observed quadratic component of the electro-optic effect $\varphi(t) \propto E^2(t)$ implies a small transient anisotropy of reflectivity that probably indicates a short-lived anisotropy of the temperature of the surface electronic states. We note that a similar phenomenon of the short-lived anisotropy of the Dirac electron gas was reported for almost intrinsic graphene \cite{König-Otto}. In that case, it was associated with relatively slow noncollinear electron-electron scattering of Dirac electrons that delays full thermalization and isotropization of the electron distribution. We can assume that surface Dirac electrons in Bi$_2$Se$_3$ demonstrate similar dynamics. In this case, the characteristic noncollinear scattering time should be $\sim$ 100 fs or less (1/8 of the period of the terahertz carrier wave) for the quadratic waveform to be clearly resolved and at the same time long enough for the effect to be detectable.

It should be noted that a specific mechanism of the quadratic electro-optic effect was recently suggested for bilayer graphene \cite{Ma}. It is caused by a higher-order nonlinearity of the band dispersion and can be relevant in our case, since in Bi$_2$Se$_3$ deviations from the linear dispersion as well as the hexagonal warping can become important for the typical Fermi level position of $\sim$ 0.3 eV from the Dirac point.

In order to identify electronic processes responsible for the linear component of the $\varphi(t)$ signal, it is instructive to compare the linear electro-optic effect to its reverse --- the linear photogalvanic effect (LPGE). For crystals with C$_{3v}$ symmetry the tensor associated with the latter is
\begin{equation}
\bordermatrix{     
      & xx & yy & zz & yz & zx & xy \cr
x &  0 & 0 & 0 & 0 & a & -b  \cr
y &  -b &  b & 0 & a & 0 & 0  \cr
z &  c &  c & d & 0 & 0 & 0
},
\end{equation}
where $x$, $y$, and $z$ are the directions of photocurrents and $xx$, $yy$, etc. are the directions of optical fields. It can be seen that this tensor has the same structure as the transposed Pockels tensor (cf. Eq. 4), and, therefore, the coefficients $a$, $b$, $c$, and $d$ determine the particular “reciprocal” effects. Since in our case the terahertz electric field acts along the sample surface, only the coefficients $a$ and $b$ are relevant in our layout.

In previous studies of ultrafast photocurrents in Bi$_2$Se$_3$, the coefficient $a$ from Eq. 13 could be formally associated with generation of the drift current induced via acceleration of photogenerated carriers by the space charge field perpendicular to the crystal surface (see e. g. Ref. \cite{Braun}). However, the processes involving electron diffusion in nonuniform electric fields are unrelated to LPGE in its classical definition \cite{Sturman}. There is no direct analogy of drift current for the electro-optic effect in Bi$_2$Se$_3$, and even if such a phenomenon existed, it would be unlikely affected by indium doping.

The coefficient $b$ is more specific and causes the $\sin3\Phi$ and $\cos3\Phi$ harmonics in the measured signal in accordance with the surface symmetry of the Bi$_2$Se$_3$ crystal. In the experiments, in which LPGE in Bi$_2$Se$_3$ was studied, the corresponding signal with three-fold symmetry was typically associated with the shift current \cite{Braun}. The latter is generated upon light-induced transitions, for which the center of mass of the electron density is shifted in the excited state relative to the ground state. The reverse effect would be virtually indistinguishable from the standard Pockels effect observed in dielectrics and caused by the distortion of electronic wavefunctions in the applied electric field. In our case, the linear electro-optic signal caused by the peak electric field of the terahertz pulse in Bi$_2$Se$_3$ is $\sim$ 5 times weaker than the signal that would be measured using a hypothetical ZnTe crystal of $\sim$ 2 nm thickness (“thickness” of the surface Dirac states \cite{Zhang}). Therefore, such instantaneous Pockels effect, caused mainly by bound electrons, cannot be immediately excluded. However, such an interpretation contradicts our conclusions made above on the probable intraband origin of the terahertz electro-optic effect in Bi$_2$Se$_3$.

In the framework of the intraband scenario, the terahertz electric field in combination with the trigonal symmetry of the system creates nonsymmetric momentum distribution of the surface electronic states (one example of such a process is the skew scattering \cite{Du}). The latter can give rise to the electro-optic effect probably involving optical transitions to or from the surface states in the vicinity of the Fermi level. This effect can be caused jointly by acceleration and anisotropic heating of the electron gas. Therefore, linear and quadratic components of the corresponding $\varphi(t)$ signal can be represented as $\varphi^{(1)}(t)\propto\int_{-\infty}^{t}E(t')e^{-(t-t')/\tau_1}dt'$ and $\varphi^{(2)}(t)\propto\int_{-\infty}^{t}E^2(t')e^{-(t-t')/\tau_2}dt'$, respectively, in the framework of a simplified model describing the time integrated response with exponential decay. If the characteristic relaxation times $\tau_1$ and $\tau_2$ are short enough with respect to the characteristic period of the terahertz electric field, the shapes of the linear and quadratic waveforms $\varphi^{(1)}(t)$ and $\varphi^{(2)}(t)$ will be very close to $E(t)$ and $E^2(t)$, respectively. In this case $\varphi^{(1)}(t)$ and $\varphi^{(2)}(t)$ will be synchronous, while the total signal -- seemingly instantaneous with respect to the terahertz field, with a very small delay determined by the values of $\tau_1$ and $\tau_2$. Such a property justifies the fit $\varphi(t) = A_1E(t) + A_2E^2(t)$ performed above.

Under our experimental conditions it is rather difficult to determine with sufficient precision the absolute value of the delay between the $\varphi(t)$ signal and the terahertz $E(t)$ waveform measured using electro-optic detection. Instead, we extract the profile of the $E^2(t)$ waveform from the transient changes of reflectivity. In order to obtain the $\Delta R/R$ trace without modifying the optical paths of the pump and probe pulses, we calculated the experimental signal as $\tilde{S}=1-I^*_y/I_y$ instead of $S=1-(I^*_y/I^*_x)/(I_y/I_x)$. Since $S$, which is defined by the polarization rotation of the reflected probe pulse, is order of magnitude smaller than $\tilde{S}$, it is possible to put $\tilde{S}\approx-\Delta R/R$. After that, modelling the transient reflectivity as $\Delta R/R\propto\int_{-\infty}^{t}E^2(t')e^{-(t-t')/\tau}dt'$ we extracted the squared terahertz field as $E^2(t)\propto(\Delta R/R)'+\Delta R/R\tau$, where $\tau$ is the decay time of the transient reflectivity.

\begin{figure}
\begin{center}
\includegraphics{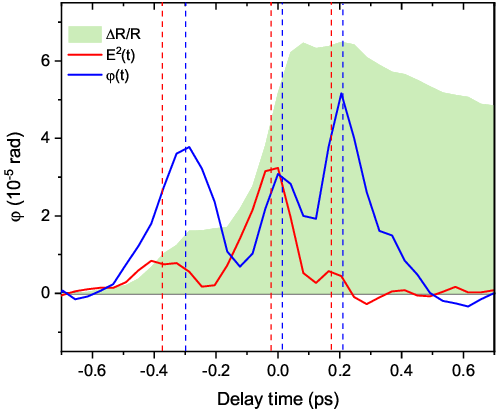}
\end{center}
\caption{Polarization rotation $\varphi(t)$ of the probe pulse induced by the terahertz pulse for the (Bi$_{1-x}$In$_x$)$_2$Se$_3$ crystal with $x$ = 0.025 (blue solid line). Transient changes of reflectivity of the same crystal induced by the terahertz pulse (green area, in arbitrary units). Squared electric field of the terahertz pulse $E^2(t)$ extracted from this $\Delta R/R$ trace is shown in arbitrary units by the red solid line (see text). Vertical dashed lines indicate positions of the three peaks of $\varphi(t)$ and $E^2(t)$ waveforms obtained by fitting 4-5 points of the curve in the vicinity of each peak by a parabola.}
\end{figure}

The obtained squared field $E^2(t)$ is shown in Fig. 10 along with the $\varphi(t)$ signal. We have found that the delay between the terahertz field and the $\varphi(t)$ waveform varies from $\sim$ 75 fs for the first peak to $\sim$ 35 fs for the second and third peaks. This variation can be due to heating of the electrons by the terahertz pulse and is another probable reason for the small temporal compression of the $\varphi(t)$ signal as compared to the $E(t)$ field measured using electro-optic detection. Such a small value of the delay implies equally small characteristic relaxation times of the nonequilibrium electron distribution that causes the $\varphi(t)$ signal and allows fitting the measured optical response as $\varphi(t) = A_1E(t) + A_2E^2(t)$.

\section{Conclusion}

In conclusion, we have discovered an electro-optic effect in the topological insulator Bi$_2$Se$_3$, induced by intense terahertz pulses. The effect was detected as rotation of the polarization of femtosecond pulses reflected from the crystal under the action of the terahertz electric field. The associated experimental signal contained linear and quadratic components, which followed the field almost instantaneously, with a delay of several tens of femtoseconds. The linear component was characterized by the three-fold symmetry with respect to crystal orientation, while the quadratic one was almost isotropic. Both components could be quenched via transition to the topologically trivial phase induced by indium doping or by means of a femtosecond laser pre-pulse that heated the electron ensemble. We ascribed the observed electro-optic effect to the transient anisotropic or nonsymmetric electron momentum distribution created via specific processes of intraband dynamics of surface Dirac electrons driven by the terahertz electric field. Our results suggest that the electro-optic response of surface topological states is considerably stronger than that of the trivial electronic states of the same material (potentially, orders of magnitude stronger). Therefore, observation of the terahertz electro-optic effect can be a sensitive probe of the non-trivial surface electronic states and serve as an alternative method of investigation of the ultrafast electron dynamics in topological materials. The functional form of the detected electro-optic response and its delay relative to the driving terahertz field contain information on the characteristic relaxation times of the anisotropic electronic distribution and on the timescales of related scattering processes. We expect that electro-optic effects of this type can be readily detected using intense terahertz pulses if a studied crystal has topological metallic surface states, while its bulk is centrosymmetric and lacks pronounced structural response. It would be interesting also to study evolution and possible enhancement of the terahertz electro-optic effect at interfaces and in heterostructures composed of topological materials.

\begin{acknowledgments}

The reported study was funded by the Russian Science Foundation, Project No. 23-22-00387.

\end{acknowledgments}

\end{document}